# Magnetic Proximity Effect as a Pathway to Spintronic Applications of Topological Insulators

*Ivana Vobornik[§,\*], Unnikrishnan Manju[†], Jun Fujii[§], Francesco Borgatti[‡], Piero Torelli[§], Damjan Krizmancic[§], Yew San Hor[#], Robert J. Cava[#], and Giancarlo Panaccione[§]*

§ Istituto Officina dei Materiali (IOM)-CNR, Laboratorio TASC, in Area Science Park, S.S.14, Km 163.5, I-34149 Trieste, Italy, † International Centre for Theoretical Physics (ICTP), Strada Costiera 11, I-34100 Trieste, Italy, ‡ ISMN-CNR, via Gobetti 101, I-40129 Bologna, Italy, #Department of Chemistry, Princeton University, Princeton, New Jersey, 08544 USA.

*To whom correspondence should be addressed: e-mail, ivana.vobornik@elettra.trieste.it; phone +39 040 3758411

ABSTRACT Spin-based electronics in topological insulators (TIs) is favored by the long spin coherence[1,2] and consequently fault-tolerant information storage. Magnetically doped TIs are ferromagnetic up to 13 K,[3] well below any practical operating condition. Here we demonstrate that the long range ferromagnetism at ambient temperature can be induced in $Bi_{2-x}Mn_xTe_3$ by the magnetic proximity effect through deposited Fe overlayer. This result opens a new path to interface-controlled ferromagnetism in TI-based spintronic devices.

MANUSCRIPT TEXT The recently discovered family of Topological Insulators (TIs) has led to flourishing of novel and fascinating physics. The simplicity of their surface states, the robustness of both topological properties and surface metallicity under external perturbations, and the prediction of novel quantized states arising from the peculiar coupling between magnetic and electric fields make TIs a perfect venue for the next generation electronic and spintronic devices.[4-7]

TIs can be charge-doped and therefore the position of the Fermi level with respect to the Dirac point can be controlled, allowing for topological transport.[8] Simultaneous magnetic and charge doping forms a gap at the Dirac point producing massive fermions at the surface.[9] TIs may also host superconductivity[10] ($Cu_xBi_{2-x}Se_3$) or ferromagnetism[3] ($Bi_{2-x}Mn_xTe_3$) via dilution of 3$d$-metal dopants.

Still, for possible device-like applications, a number of key problems remains to be resolved, both from the conduction and magnetic point of view. These are in particular regarding disentanglement and independent control of the bulk and surface properties of a TI.[11] Very recently it was shown that the increased surface to bulk ratio in the TI nano-ribbons leads to realization of the topological transport in a device-like manner.[12] Concerning the magnetic properties, TIs like $Bi_{2-x}Mn_xTe_3$ do display long range ferromagnetism, but with a Curie temperature one order of magnitude lower than needed for applications.[3] We succeeded in controlling magnetic properties of $Bi_{2-x}Mn_xTe_3$ by exploiting magnetic proximity effect: thin ferromagnetic overlayers (in our case Fe) drive the TI ferromagnetic at room temperature, opening new frontiers in exploitation of magnetically controlled interfaces with topological transport in new spintronic devices.

The Mn-doped $Bi_2Te_3$ crystals studied here are from the same sample batches reported in reference 3, with nominal formula $Bi_{1.91}Mn_{0.09}Te_3$. Scanning Tunneling Microscopy (STM) images after *in situ* room temperature cleavage are shown in Figure 1. In agreement with previous low temperature results,[3] substitutional Mn atoms are visible as triangular shadows in a triangular lattice of Te atoms.

Quantitative analysis of the STM topography indicates that the Mn content exposed at the surface corresponds to x = 0.035, rather than nominal x = 0.09, suggesting a higher dilution of Mn in the surface region. The distribution of Mn atoms does not show any clustering effect, confirming previous results.[3]

Recently the response of the topological insulators to magnetic perturbations has been studied.[13] The authors of reference 13 (including one of the coauthors of the present work (R.J.C.)) progressively evaporated Fe on the topological insulator, and followed the evolution of the band structure by angle-resolved photoemission (ARPES). The response of the surface was surprising: Fe, being the magnetic impurity, did have the expected impact of opening a gap at the original Dirac point, but moreover, the layer just below the interfacial one was electrically reorganized, so that the new Dirac cones appeared.[13] These new states did remain gapless, while the total number of cones remained odd,[13] as a prerequisite for the topologically protected surface.[1] It was possible to monitor the band structure evolution for the surfaces with up to 0.5 ML of Fe, which by itself is a surprising result for the case of strongly surface-sensitive technique such as ARPES: not only the surface states were preserved even in the presence of Fe overlayer (as typically never happens with ordinary surface states[14]), but they even gained in sharpness at an intermediate coverage of ~0.3 ML. Therefore the topological properties at the interface with Fe do remain preserved and even emphasized. This is also the reason why we chose Fe for our study instead of some other magnetic material.

The growth of Fe overlayers was monitored by Low Energy Electron Diffraction (LEED) and STM in the thickness range of 0-10 Å. The LEED pattern progressively fades away after the deposition of Fe indicating the absence of structural long range order in the Fe overlayer. High resolution STM topography images are displayed in Figure 1a-h. At very low coverage (0.1-0.5 Angstroms, Figure 1c-f) the random distribution of Mn atoms (black triangles) and of the Fe islands (white spots) is indicative of the absence of a preferential interaction between the adsorbed Fe and Mn in $Bi_{1.91}Mn_{0.09}Te_3$. Several sizes of Fe islands in the deposited overlayer are found: faint spots that represent very small clusters, and small and large islands with apparent diameters and heights ranging from 2 to 10 Å and 1 to 6 Å, respectively. The variation of the apparent height of the Fe islands at low coverage suggests the presence of an important variation of the electronic density, as often observed when metallic structures are deposited onto semiconducting or insulating substrates (and vice-versa).[15] Increased Fe deposition reduces the number of faint spots and most Fe atoms form 6 Å high islands. At 10 Å coverage, the Fe islands are homogeneously distributed with very similar sizes (Figure 1g). At this coverage, long range ferromagnetism is present in the overlayer up to room temperature, as demonstrated by the X-ray Magnetic Circular Dichroism (XMCD) signal measured at the Fe edge (Figure 3).

We present in Figure 2 the X-ray Absorption Spectroscopy (XAS) data from $Bi_{1.91}Mn_{0.09}Te_3$. The acquisition was done in Total Electron Yield (TEY) mode[16] in order to enhance the near surface-interface sensitivity.[17] The double peak structure of the $L_2$ edge near 650 eV photon energy and the smooth shoulder at ~640 eV in the vicinity of the $L_3$ edge are the signatures of a dilute Mn systems, with no trace of oxidation and/or contamination.[18,19] We compare in Figure 2 the experimental XAS spectra with the theoretical one, obtained considering the electric-dipole allowed transitions between the Mn $3d^5$ ground-state and final $2p^53d^6$ configurations in spherical symmetry, i.e. without crystal field effects (see Supporting Information for calculation details). The smearing of the multiplet features in the experimental results indicates that Mn bears a mixed-valence state with itinerant character, although the ground state is unambiguously dominated by the $3d^5$ electronic configuration. The dominant $3d^5$ state does not significantly modify the ground-state symmetry, explaining the absence of distinct charge-transfer satellites in XAS spectra.[20,21] The presence of a mixed valence state suggests a rearrangement of the local DOS associated with changes in the Mn environment. This electronic configuration has interesting analogies to half-metallic systems and dilute magnetic semiconductors, where doped Mn atoms have localized magnetic moments but delocalized $3d$ electrons, yielding long range FM properties.[18,19] As seen from Figure 2, the Mn line shape before and after deposition remains virtually identical, thus excluding any alloying between the Fe in the overlayer and the Mn in the $Bi_{1.91}Mn_{0.09}Te_3$ crystal.

Having ascertained the stable character of Mn as a doped impurity before and after the Fe deposition, we now address the magnetic properties of the interface. Figure 3 presents the magnetization dependent $L_{2,3}$ spectra of Fe and Mn measured for 10 Å Fe/$Bi_{1.91}Mn_{0.09}Te_3$, at 22 K, i.e. already above the bulk $T_C$ of 13 K. A clear XMCD signal, showing the presence of long range ferromagnetic ordering, is observed for both Fe and Mn. The correlated "up-down" feature of the XMCD for Mn and Fe shows that there is an antiparallel alignment of Mn and Fe magnetizations. This behavior is strikingly similar to what was observed in other dilute Mn-based systems.[18] We note that the observed Mn XMCD signal is about one order of magnitude smaller than the calculated one (Figure 3b). Assuming a uniform distribution of ferromagnetic Mn atoms in the bulk TI over the thickness typically probed by XAS,[17] our result suggests that only the topmost layers of the $Bi_{1.91}Mn_{0.09}Te_3$ are ferromagnetic.

The temperature dependence of the magnetic signal is summarized in Figure 4, where the element selective XMCD hysteresis loops are displayed. The loops have opposite polarity due to the antiparallel alignment of the Mn and Fe long range magnetizations. The evolution of the magnetic signal from Mn follows that of Fe - they have the same coercive field and increasing magnetic signal with decreasing temperature - thus confirming a robust magnetic Mn-Fe coupling. The shape of the Fe loops is not completely square, suggesting the presence of both small and large magnetic domains (in agreement with the STM characterization of the Fe overlayer) and not fully saturated magnetic regime. We observe an increase of the coercive field with decreasing temperature. Such behaviour is frequently observed in low dimensional ferromagnetic systems and attributed to the reduction of the thermal fluctuations that help overcoming the energies of nucleation and motion of the domain walls.[22] The Mn magnetization, although significantly reduced, remains finite even at room temperature, while the Fe magnetization is sizably larger. This supports our conclusion that there is no alloying between Mn and Fe since in an alloy both magnetizations would vanish at the same temperature. In our case Mn and Fe long range magnetizations are qualitatively coupled but quantitatively different.

The fact that the magnetization is not fully saturated should also be taken into account: only a fraction of the Mn atoms in the bulk would be magnetically polarized, and we argue that those atoms are in the layers close to the Fe overlayer. It is important to emphasize that in the full temperature range explored (22 K to 300 K), a magnetic XMCD signal at the Mn $L_{2,3}$ edge is detected only in the presence of the Fe overlayer. This indicates that the magnetic behavior above the bulk $T_C$ of 13 K is only due to the proximity effect of the Fe film. The XMCD results reveal that the long range magnetization lies in the surface plane; thus the Mn magnetization at the interface is magnetically independent of the intrinsic bulk magnetization of $Bi_{2-x}Mn_xTe_3$ (at T<13 K), which is perpendicular to the plane. This result is consistent with theoretical predictions for the FM order on the surface of topological insulators induced by the spin-spin interaction mediated by the surface states.[23]

Our results clearly demonstrate that the magnetic ordering can be maintained at the surface of a TI under conditions suitable for applications. Recent findings indicate that the surface of a TI maintains its topological character in the presence of external magnetic perturbations (Fe overlayers) [reference]. In this view, the simple concept of FM/TI interface designates the route of exploiting magnetism in conjunction with topological transport. TI-based devices may be engineered and controlled by changing the structure and the local environment of the TI surfaces through the usage of deposited overlayers. To this end further work to explore the properties of the topological surface states in the presence of the magnetic (and nonmagnetic) interfaces and their dependence on band filling will be of significant interest.

ACKNOWLEDGMENT This work was partially funded by CNR-IOM. The work at Princeton was supported by the NSF MRSEC program, grant DMR-0819860.

SUPPORTING INFORMATION AVAILABLE  Additional details on $Bi_{2-x}Mn_xTe_3$ single crystals, STM measurements, XMCD measurements, calculations. This material is available free of charge via the Internet at http://pubs.acs.org.

FIGURE CAPTIONS

FIGURE 1. STM Characterization of the clean and Fe covered $Bi_{1.91}Mn_{0.09}Te_3$. Left 100x100 nm2 STM topography images; right, finer scale images: (a) as cleaved $Bi_{1.91}Mn_{0.09}Te_3$ (+10 mV, 3.8 nA), (c) after 0.1 Å Fe deposition (+19 mV, 1.8 nA), (e) after 0.5 Å Fe deposition (+680 mV, 0.3 nA). Right. atomically resolved STM images (20x20 nm2) (b) as cleaved $Bi_{1.91}Mn_{0.09}Te_3$ (-20 mV, 3.8 nA), (d) 0.1 Å Fe deposition (+110 mV, 1.0 nA), (f) (30x30 nm2) 0.5 Å Fe deposition (+620 mV, 0.4 nA). In a) and b) dark areas (triangles) correspond to Mn atoms. In c)-d)-e)-f) Fe atoms (white spots) are randomly distributed (no preferential Mn-site is observed), with different island sizes. At 0.5 Å coverage (e,f), both the height and size of the islands converge to a constant value. At 10 Å coverage (g, h), rough coalescence of the Fe islands (0.3 nm in average size) is found, with no long range order. Above 6 Å of Fe thickness, a remnant magnetic signal in the Fe film is detected at room temperature.

FIGURE 2. Mn $L_{2,3}$ XAS experiment and calculation. Lineshape comparison of the Mn $L_{2,3}$ XAS spectra measured at room temperature before (blue) and after (red) the deposition of the Fe thin film (10 Å) overlayer. The spectra were obtained by averaging the TEY signals of the magnetization dependent Mn spectra measured under remnant magnetization conditions, and normalized to the maximum of their intensities. The experimental results are compared to theoretical calculations (black) of the XAS spectrum for a Mn $d^5$ ground state configuration in O(3) symmetry (see supporting information for details).

FIGURE 3. Fe and Mn $L_{2,3}$ XMCD experiment and calculation. XAS/XMCD results for the Fe/$Bi_{1.91}Mn_{0.09}Te_3$ interface (T= 22 K, Fe thickness 10 Å). Spectra have been shifted vertically for sake of comparison. Inset: experimental geometry of the XAS/XMCD experiment. The magnetic field is applied along the surface plane. XMCD measures the differences at the $L_3$ and $L_2$ edges when reversing the magnetization (M+ and M- in the figure) at fixed photon helicity. Synchrotron Radiation impinges onto the sample with an angle of 45 degrees with respect to the surface normal. Iron: a) Magnetization dependent Fe $L_{2,3}$ XAS/XMCD spectra. Manganese: b) Background subtracted magnetization dependent Mn $L_{2,3}$ spectra (red and black curves) are compared to theoretical calculations (continuous grey line) for a Mn $d^5$ ground state configuration in O(3) symmetry and an internal exchange field of 10 meV. The experimental XMCD data (blue line), amplified by a factor 10, is in good agreement with the theoretical calculation (pink curve). Comparison of (a) and (b) shows that the "up-down" features of the XMCD curves are opposite for Mn and Fe, indicating antiparallel alignment of the long range Mn and Fe magnetizations.

FIGURE 4. Mn and Fe XMCD hysteresis loops vs. temperature. Element selective XMCD hysteresis loops for Fe (black) and Mn (red). Hysteresis curves are measured with the photon energy fixed at the maximum of the XMCD magnetic signal ($L_3$), and sweeping the magnetic field. Normalization is obtained through on-edge and off-edge measurements. The orientation of the Mn loops is opposite to that of the Fe loops, dictated by the antiparallel alignment of the magnetizations. The experimental geometry probes only the in-plane magnetization of both Fe and Mn. Both the magnetic signal (height of the loops) and the coercive field (width of the loops) increase when temperature is decreased, suggesting a robust magnetic coupling. The Mn XMCD signal at room temperature, although clearly visible, is close to zero, while Fe magnetic signal, as expected for a 1 nm ferromagnetic film, is still large.

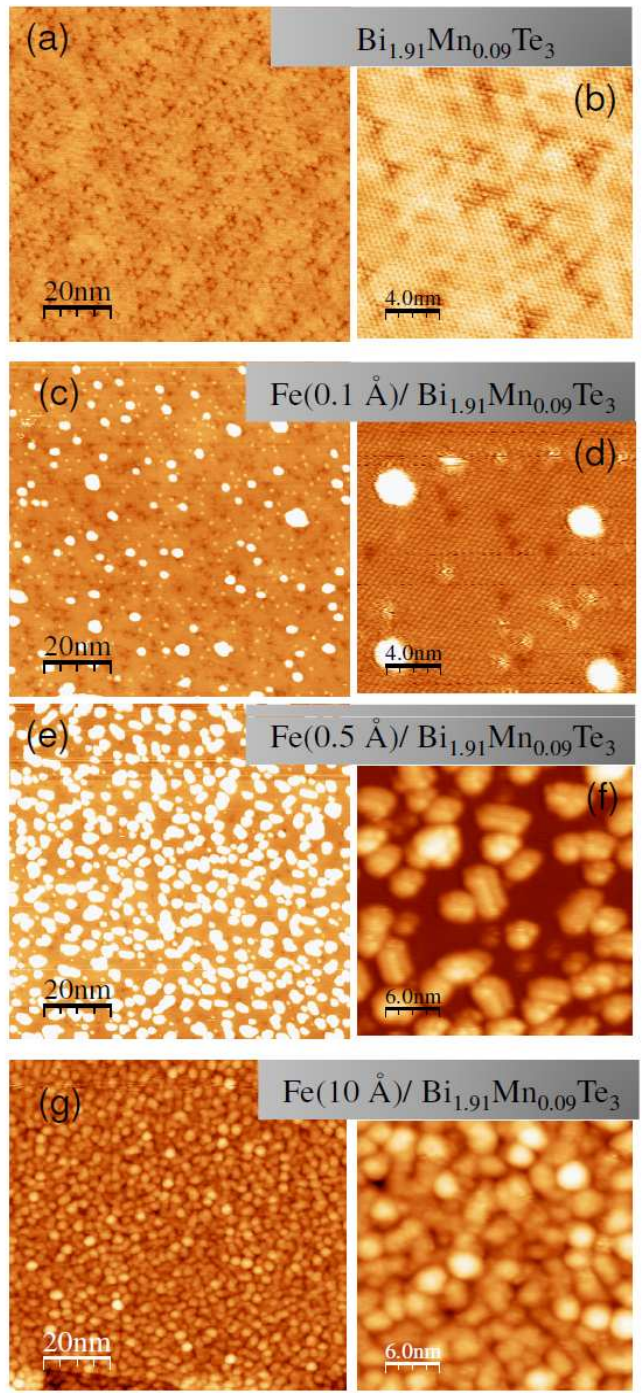

**FIGURE 1**

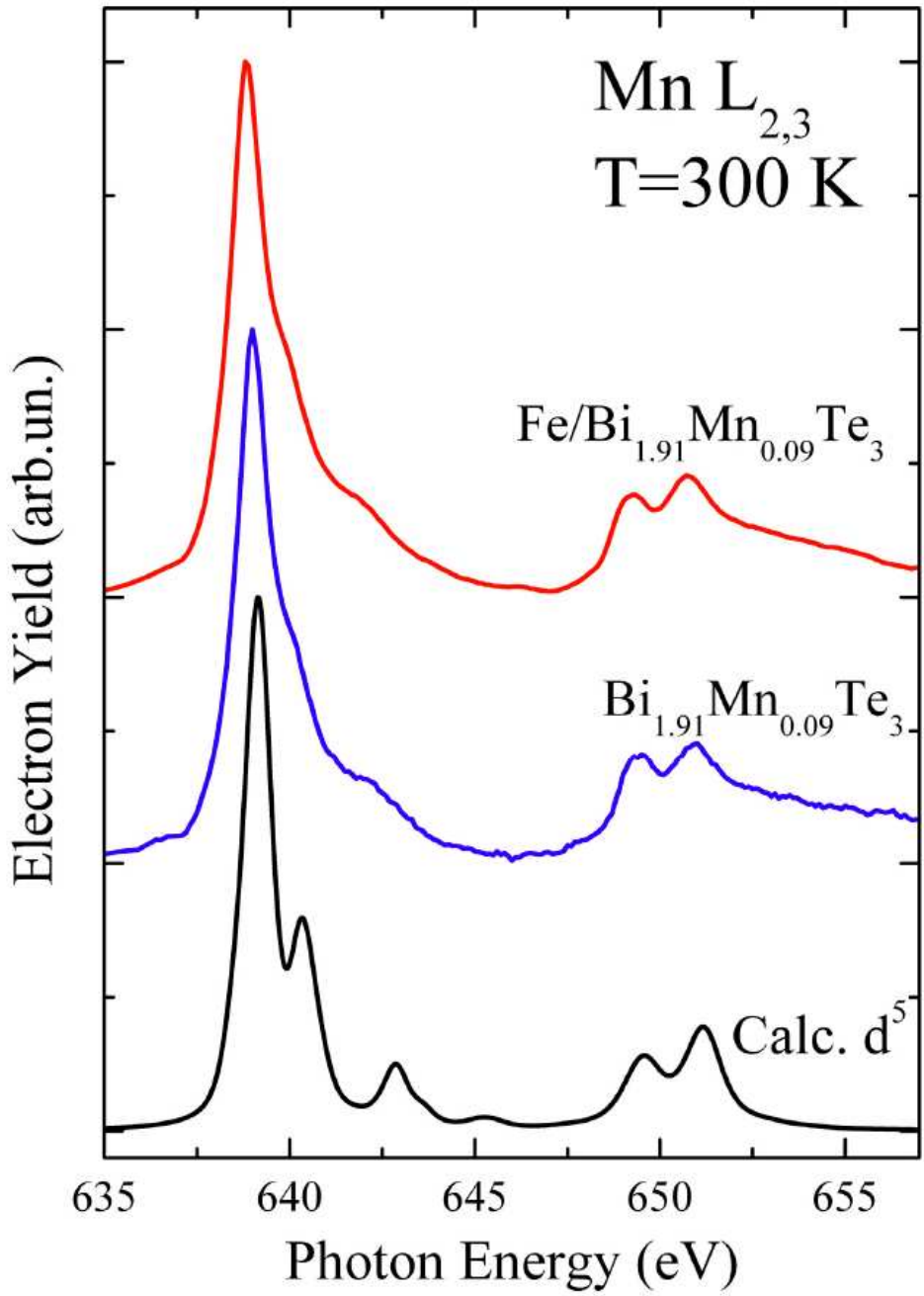

**FIGURE 2**

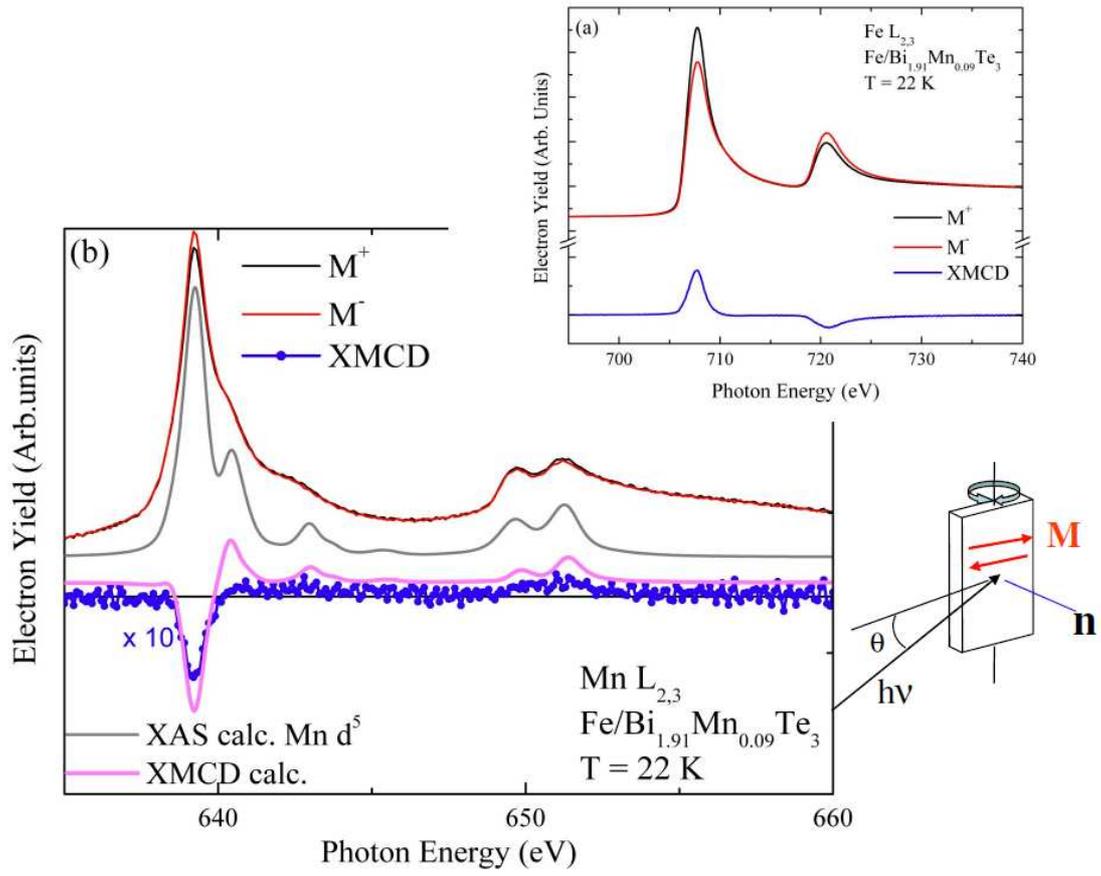

**FIGURE 3**

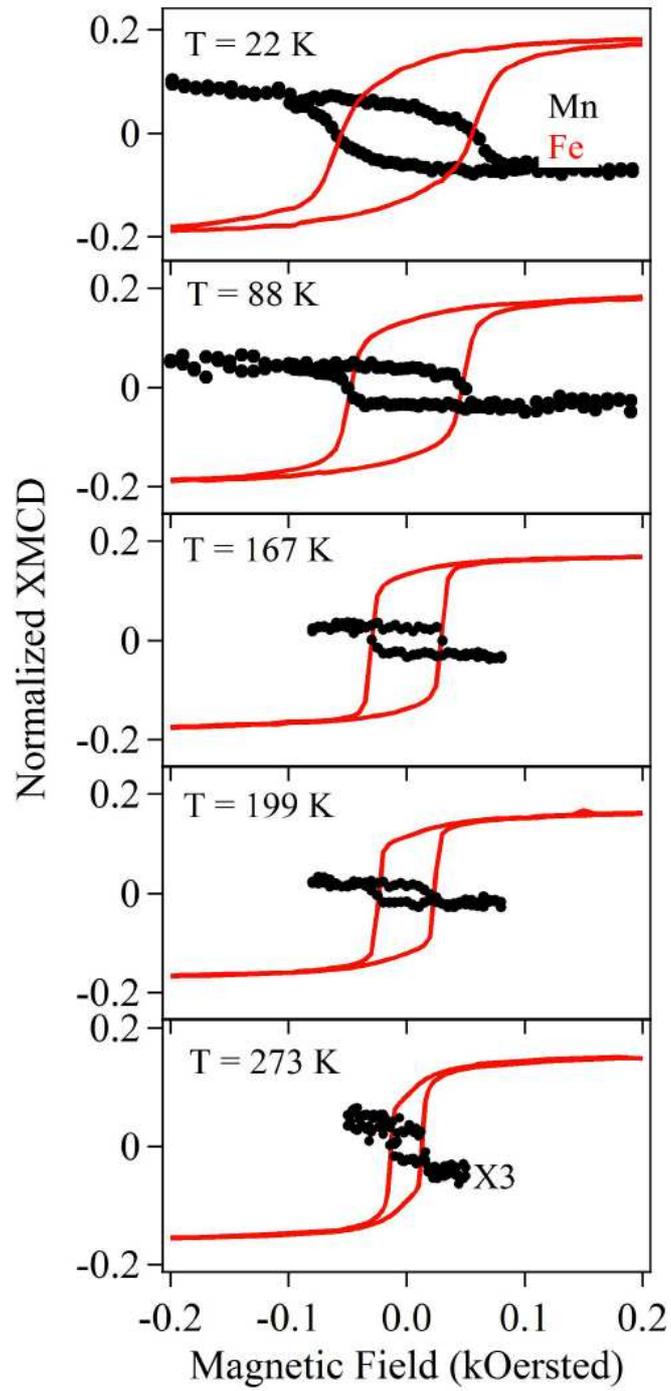

**FIGURE 4**

**Supporting information:**

**Sample preparation and characterization:** High purity elemental Bi (99.999%), Mn (99.99%), and Te (99.999%) were used for the $Bi_{2-x}Mn_xTe_3$ crystal growth, employing a nominal x value of 0.1. A modified Bridgeman crystal growth method was employed. The crystal growth for $Bi_{2-x}Mn_xTe_3$ involved cooling from 950 to 550 °C over a period of 24 h and then annealing at 550 °C for 3 days; silver-colored single crystals were obtained. The crystals were confirmed to be single phase and identified as having the rhombohedral $Bi_2Te_3$ crystal structure by X-ray diffraction. The Mn concentration in the bulk crystals was determined by the ICP-AES method and was very close to the nominal compositions (the analytically determined composition is used in this report). Details of the structural and magnetic characterization as a function of temperature, magnetic field and doping can be found in ref 1. Clean surfaces of $Bi_2Te_3$ and $Bi_{1.91}Mn_{0.09}Te_3$ were obtained by cleaving their respective crystals *in-situ* (pressure better than $8 \times 10^{-11}$ mbar) at 22 K. Fe was deposited by e-beam evaporation technique with the substrate crystal kept at room temperature. The deposition rate was monitored using a quartz microbalance and was calibrated to be $\leq 0.1$ Å/min. During the whole process, the vacuum inside the chamber was better than $2 \times 10^{-10}$ mbar. The samples were then transferred in UHV conditions from the STM/Fe growth chamber to the end station for the XMCD measurements and vice-versa.

**STM measurements:** STM measurements were performed using a home-built room temperature UHV STM[2]. A chemically etched W wire was used for a tip. The STM topographic images were acquired with a constant current mode. The sample bias voltage and the tunneling current for each image are indicated in the figure captions.

**X-ray techniques and experiments:** X-ray absorption spectroscopy (XAS) and X-ray Magnetic circular Dichroism (XMCD) combine sensitivity to magnetic properties with chemical specificity and surface sensitivity. In present experiment, circularly polarized photons were swept across the $L_{2,3}$ absorption edges of Fe and Mn. The differences in the measured intensities at the $L_3$ and $L_2$ edges for parallel and antiparallel orientation of photon helicity and magnetization directions are quantitatively related by sum rules to the size, the orientation and the anisotropies of the spin and orbital magnetic moments. A comprehensive review of principles and application of XMCD can be found in ref 3. We acquired the XAS and XMCD spectra at the APE beamline[2] of the Elettra Synchrotron (Trieste, Italy) in the temperature range of 20-300 K and in a base vacuum $< 2 \times 10^{-10}$ mbar. All XAS and XMCD measurements were obtained by measuring the sample drain current - Total Electron Yield (TEY) - with an energy resolution of 150 meV and a circular polarization rate of 70%. The photon incidence angle was kept fixed at 45° from the sample surface normal. Beam spot size on the sample surface was 100x200 µm$^2$. The samples were magnetized *in situ* by applying a magnetic field up to a maximum of 2000 Oe in pulsed mode parallel to the sample surface plane. XMCD spectra were obtained in remnance (i.e. zero applied field) by reversing the magnetization at fixed polarization; reference measurements were obtained with fixed magnetization by reversing the polarization from left to right circular resulted in identical spectra, thus excluding experimental artifacts in the XMCD signals. All spectra were normalized by the intensity of the incident beam, which is given by the TEY of a gold mesh far enough from the magnet to avoid any effect on its electron yield.

**XAS Background Subtraction** The comparison between experimental and theoretical lineshapes of Mn $L_{2,3}$ for both XAS and XMCD was obtained after background subtraction. The raw Mn $L_{2,3}$ XAS spectra displayed at the top of the figure are first normalized to the incoming photon flux measured contemporarily on a mesh. The procedure is described in Fig. S3: a straight baseline fitting the slope of the raw TEY signal and matching the intensity of the $L_3$ pre-edge region has been subtracted to the data. To avoid artifacts, the same baseline was used for each set of magnetization dependent spectra. If the background of the magnetization dependent spectra did not match perfectly, the baseline was shifted vertically to compensate the difference in the background.

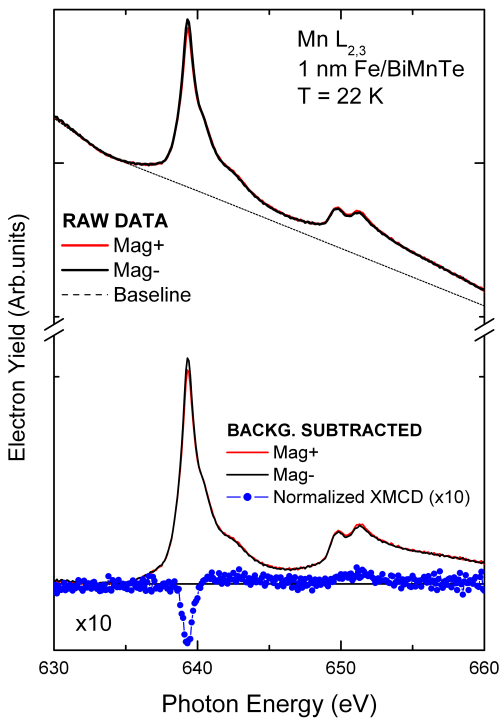

FIGURE S1. Details of the background subtraction procedure. Top spectra (black and red curve): raw data. Dashed line: baseline subtracted. Bottom spectra: magnetization dependent spectra (black and red curve) after background subtraction, blue filled circles, experimental XMCD.

**Calculation details:** Atomic multiplet calculations were performed at intermediate coupling using the codes developed by R. D. Cowan and co-workers[4] and modified by B.T. Thole[5]. The Mn $L_{2,3}$ XAS spectra were obtained from the electric-dipole-allowed transitions between the ground-state $3d^5$ and the final-state $2p^53d^6$ configurations. Interatomic screening and mixing was taken into account by reducing the atomic values of the Slater integrals $F^k(3d,3d)$, $F^k(2p,3d)$, and $G^k(2p,3d)$ to 70% of their values. An exchange field of $g\mu BH$=10 meV was used. The XMCD curve is the difference between calculated absorption spectra obtained from the sum of all possible transitions for an electron excited from the *2p* level into a *3d* level with right/left circularly polarized radiation. The theoretical XMCD for a pure Mn $d^5$ ground state was obtained in O(3) spherical symmetry. The calculated spectra include a Lorentzian of 0.25 (0.4) eV for the $L_3(L_2)$ edge to account for intrinsic linewidth broadening and a Gaussian of σ= 0.1 eV for instrumental broadening.